\documentclass[twocolumn,showpacs,preprintnumbers]{revtex4}
\usepackage{amssymb}
\usepackage{amsfonts}
\usepackage{amsmath}
\usepackage{graphicx}
\usepackage{dcolumn}
\usepackage{bm}

\input{tcilatex}

\begin{document}

\title{On \ the \ statistical description \ of \ classical \ open \ systems
\ with \ integer variables by the Lindblad equation}
\author{E. D. Vol}
\email{vol@ilt.kharkov.ua}
\affiliation{B. Verkin Institute for Low Temperature Physics and Engineering of the
National Academy of Sciences of Ukraine 47, Lenin Ave., Kharkov 61103,
Ukraine.}
\date{\today }

\begin{abstract}
We propose the consistent \ statistical approach to consider a wide class of
classical open systems whose states are specified by a set of positive
integers(occupation numbers).Such systems are often encountered in \
physics, chemistry, ecology, economics and other sciences.Our statistical
method based on ideas of quantum theory of open systems takes into account
both discreteness of the system variables and their time fluctuations - two
effects which are ignored in usual mean field dynamical approach.The method
let one to calculate the distribution function and (or)all moments of the
system of interest at any instant.As descriptive examples illustrating the
effectiveness of the method we consider some simple models:one relating to
nonlinear mechanics,and others taken from population biology .In all this
examples the results obtained by the method for large occupation numbers
coincide with results of purely dynamical approach but for small numbers
interesting differences and new effects arise.The possible observable
effects connected with discreteness and fluctuations in such systems are
discussed.
\end{abstract}

\maketitle

\section{Introduction}

Among the vast number of classical open dynamical systems under
consideration in physics, chemistry, biology, economics and other sciences
there are many such whose states \ in accordance with the sense of the
problem are specified by \ a set of integer variables $\left( \left\{
n_{i}\right\} \right) $\ , where $i$\ = 0,1,2,..N (N-number of degrees of
freedom). For example in physics $n_{i}$-\ are occupation numbers of cell
states in phase space, in chemistry - numbers of molecules of reactive
elements, in ecology -numbers of individuals in populations which live in
the area and interact with other populations, in economics the number of
companies\ operating on the market. Since all these systems are classical
their dynamics as a rule is described by a system of differential equations
of the form%
\begin{equation}
\frac{dn_{i}}{dt}=F_{i}\left( \left\{ n_{\alpha }\right\} \right) ,
\label{q1}
\end{equation}%
where $F_{i}(\left\{ n_{\alpha }\right\} )$ -some nonlinear functions,
depending on concrete problem. Obviously notation Eq. (\ref{q1}) implies
that variables $n_{i}$ in system Eq. (\ref{q1}) -are considered as
continuous. In the case when all $n_{i}\gg 1$ such approach can be easy
justified. From the physical point of view the system of equations (\ref{q1}%
) corresponds to mean field approximation and $n_{i}$ are occupation numbers
averaged over some appropriate statistical ensemble. In the case when all $%
n_{i}\gtrsim 1$, dynamical description becomes inadequate and \ the question
naturally arises: is there consistent statistical approach which takes into
account both discreteness of variables $n_{i\text{ }}$\ and their time
fluctuations\ that may be not small.\ In addition it is naturally to demand
that such approach gave the same results as dynamical description in the
large $n_{i\text{ }}$\ limit. In this paper we propose such approach based
on the ideas of quantum theory of open systems (QTOS) and consider some
examples that demonstrate its effectiveness. The rest of the article
organized as follows. In the Sect.2 we briefly describe minimal information
from (QTOS) which is necessary for understanding of the method used and
present main steps of our method. In Sect.3 we consider simple model of
nonlinear autonomous oscillator with soft exciting mode and give its
statistical description on the basis of the method proposed. All the main
features of the method clearly come to light already in this representative
example. In Sect.4 with the help of our approach we consider some problems
from population dynamics relating to evolution of two interaction
populations living in a certain area..We show that in the case of small
populations statistical description leads to a number of differences from
the ordinary dynamical picture. On the other hand in the case of large
occupation numbers both descriptions are virtually identical. In conclusion
we discuss some generalizations of the method and its possible experimental
verification.

\section{Description of the method}

In this section we briefly remind the main points of the method proposed by
author earlier \cite{1a} which allows one to make the transition from known
dynamical equations of classical open system to the master equation for its
quantum analogue. The method based on the correspondence that can be set
between quantum master equation in the Lindblad form and the Liouville
equation for distribution function in phase space \ of the classical system
of interest. This correspondence allows one using classical equations of
motion to restore the form of all operators involved in the Lindblad
equation. Thereby we can apply the procedure of quantization at least in
semiclassical approximation in the case of a large class of nonhamiltonian
dynamical systems. In short \ (all details see in \cite{1a} ) the recipe of
quantization proposed consists of three consecutive steps.

\textbf{Step1:} The input classical dynamical equations should be presented
in the form allowed the quantization (FAQ). For the purposes of present
paper the most convenient form is complex \ representation of equations of
motion:%
\begin{equation}
\frac{dz_{i}}{dt}=-i\cdot \frac{dH}{dz_{i}^{\ast }}+\dsum\limits_{\alpha
}\left( \overline{R_{\alpha }}\frac{dR_{\alpha }}{dz_{i}^{\ast }}-R_{\alpha }%
\frac{d\overline{R_{\alpha }}}{dz_{i}^{\ast }}\right) ,  \label{q2}
\end{equation}%
where $z_{i}=\frac{x_{i}+iy_{i}}{\sqrt{2}},$\ $z_{i}^{\ast }=\frac{%
x_{i}-iy_{i}}{\sqrt{2}}$ are complex dynamical coordinates of the system of
interest, $H$, $R_{\alpha }$, $\overline{R_{\alpha }}$ are functions of $%
z_{i}$, $z_{i}^{\ast }$ ( $H$ is real function, and $R_{\alpha }$, $%
\overline{R_{\alpha }}$ are complex, $\overline{R_{\alpha }}$ means function
which conjugate to $R_{\alpha }$). It is necessary to emphasize that it is
the most delicate step of the method because it is difficult exactly to
formalize this point.

\textbf{Step2.} Having in hands representation Eq. (\ref{q2}) we can use
classical function $H$, $R_{\alpha }$, $\overline{R}_{\alpha }$ and with
their help determine their quantum analogues - operators $\widehat{H}$, $%
\widehat{R_{\alpha }}$, $\widehat{R^{+}}$. For this purpose the variables $%
z_{i}$, $z_{i}^{\ast \text{ }}$must be replaced by the correspondence Bose
operators $\widehat{a}_{i}$\ and $\widehat{a}_{i}^{+}$ with usual
commutation rules:\ $\left[ \widehat{a}_{i},\widehat{a}_{j}^{+}\right]
=\delta _{ij}$, $\left[ \widehat{a}_{i},\widehat{a}_{j}\right] =\left[ 
\widehat{a}_{i}^{+},\widehat{a}_{j}^{+}\right] =0$.

\textbf{Step3.} The operators $\widehat{H}$, $\widehat{R_{\alpha }}$, $%
\widehat{R_{\alpha }^{+}}$\ found in this manner should be substituted into
the quantum Lindblad equation for the evolution of density matrix of the
system:%
\begin{equation}
\frac{d\widehat{\rho }}{dt}=-i\left[ \widehat{H},\widehat{\rho }\right]
+\dsum\limits_{\alpha }\left[ \widehat{R_{\alpha }}\cdot \widehat{\rho },%
\widehat{R_{\alpha }^{+}}\right] +\left[ \widehat{R_{\alpha }},\widehat{\rho 
}\cdot \widehat{R_{\alpha }^{+}}\right] .  \label{q3}
\end{equation}%
The correspondence principle guarantees us that such approach will give
correct description of evolution of quantum open system at least with the
accuracy up to the first order in 
h{\hskip-.2em}\llap{\protect\rule[1.1ex]{.325em}{.1ex}}{\hskip.2em}%
.

\section{Representative model. Autonomous nonlinear oscillator with self
exciting mode}

It is convenient to demonstrate on concrete example all the features of the
approach proposed. We consider an oscillator with nonlinear damping in
situation when its equilibrium point loses stability and small fluctuations
switch the system to the new stationary state that corresponds to the closed
trajectory (limit cycle). The following system of equations gives the
correct mathematical description of the behavior of the oscillator in the
vicinity of bifurcation point (see \cite{2a}):

\begin{eqnarray}
\frac{dx}{dt} &=&\omega y+\mu x-x\left( x^{2}+y^{2}\right) ,  \label{q4} \\
\frac{dy}{dt} &=&-\omega x+\mu y-y\left( x^{2}+y^{2}\right) ,  \notag
\end{eqnarray}%
\ where $x$, $y$ are coordinates of oscillator in phase space and $\omega $
-its frequency. The system of the equations of motion Eq. (\ref{q4}) can be
written in the complex form \ as one equation%
\begin{equation}
\frac{dz}{dt}=-i\omega z+\mu z-2z\left\vert z\right\vert ^{2},  \label{q5m}
\end{equation}%
\ where $z=\frac{x+iy}{\sqrt{2}}$. It is easy to show that equation Eq. (\ref%
{q5m}) can be represented in the FAC. For this purpose we introduce the
functions $H=\omega z^{\ast }z$, $R_{1}=\sqrt{\mu }z^{\ast }$ and $%
R_{2}=z^{2}$.\ One can verify by direct checking that r.h.s. of Eq. (\ref%
{q5m}) can be rewritten in the form%
\begin{equation}
\frac{dz}{dt}=-i\frac{\partial H}{\partial z^{\ast }}+\left( \overline{R_{1}}%
\frac{\partial R_{1}}{\partial z^{\ast }}-R_{1}\frac{\partial \overline{R_{1}%
}}{\partial z^{\ast }}\right) +\left( \overline{R_{2}}\frac{\partial R_{2}}{%
\partial z^{\ast }}-R_{2}\frac{\partial \overline{R_{2}}}{\partial z^{\ast }}%
\right) .  \label{q6}
\end{equation}%
According to the recipe of quantization the Lindblad equation for quantum
analog of the system Eq. (\ref{q4}) can be written in the form%
\begin{equation}
\frac{d\rho }{dt}=-i\left[ H,\rho \right] +\left[ R_{1}\rho ,R_{1}^{+}\right]
+\left[ R_{2}\rho ,R_{2}^{+}\right] +h.c,  \label{q7}
\end{equation}%
where \ $\widehat{H}=\omega \widehat{a}^{+}\widehat{a}$, $\widehat{R}_{1}=%
\sqrt{\mu }\widehat{a}^{+}$ and $\widehat{R}_{2}=\widehat{a}^{2}$.

We are interesting in the stationary solution of \ Eq. \ref{q7} and from
physical considerations we will seek it in the form: \ $\widehat{\rho }=%
\widehat{\rho }(\widehat{N})=\sum \left\vert n\right\rangle \rho
_{n}\left\langle n\right\vert ,$ where $\left\vert n\right\rangle $ are the
eigenfunctions of the operator $\widehat{N}=\widehat{a}^{+}\widehat{a}$. It
is convenient to introduce the generating function $G(u)$ for the
coefficients $\rho _{n}$. By definition $G(u)=\sum\limits_{n=0}^{n=\infty
}\rho _{n}\cdot u^{n}.$One can obtain from Eq. \ref{q7} in stationary case
the following equation for the function $G(u)$:%
\begin{equation}
(1+u)\frac{d^{2}G}{du^{2}}-\mu u\frac{dG}{du}-\mu G(u)=0.  \label{q8}
\end{equation}%
The solution\ of Eq. \ref{q8} that satisfies all conditions of the problem
may be represented as%
\begin{equation}
G(u)=\frac{\Phi (1,\mu ,\mu (1+u))}{\Phi (1,\mu ,2\mu )},  \label{q9}
\end{equation}%
where $\Phi (a,c,x)$ is well known confluent hypergeometric function (see 
\cite{3a})$\ $- $\Phi (a,c,x)=1+\frac{ax}{c}+\frac{a(a+1)}{2!c(c+1)}x^{2}+...
$. Note that condition $G(u)=1$ corresponds to normalization of $\ \widehat{%
\rho _{st}}(n)-$ namely\ $\sum\limits_{n}\rho _{n}=1$. Having in hands the
expression Eq. \ref{q9} for the generation function one can find the average
values for any physical quantity of interest in the stationary state that is
all moments of the distribution $\rho (n)$. For example moments of first and
second order are determined by relations: $\overline{n}=\sum\limits_{n}n%
\cdot \rho _{n}=\frac{dG}{du}$, $\overline{n^{2}}-\overline{n}=\frac{d^{2}G}{%
du^{2}}$. (all derivatives are taken at point $u=1$). Let us consider now
the behavior of our system in two limiting cases: $\mu \gg 1$ and $\ \mu $ $%
\ll 1$. When $\mu \gg 1$ using the asymptotic formula for $\Phi (a,c,x)$
(see \cite{3a})we obtain for $G\left( u\right) $ the next expression%
\begin{equation}
G(u)\approx e^{\mu (u-1)}\cdot \left( \frac{1+u}{2}\right) ^{1-\mu },
\label{q10}
\end{equation}%
The average number of quanta in stationary state $\ \overline{n}$ and their
dispersion $\sigma =\overline{n^{2}}-\overline{n}^{2}$ in this case are: $%
\overline{n}=\frac{\mu +1}{2}$ and $\ \sigma =\frac{3\mu +1}{4}$. The
relative fluctuation of quanta generated in stationary state is $\frac{\sqrt{%
\sigma }}{\overline{n}}=\frac{\sqrt{3\mu +1}}{\mu +1}$ tends to zero when $%
\mu \gg 1$. The obtained result testifies validity of deterministic approach
in this case since in classical case $n_{cl}=\left\vert z\right\vert ^{2}=%
\frac{\mu }{2}$ when the system is moving along the limit cycle. Now let us
consider the opposite case when $\mu \ll 1$. Using the series expansion for $%
\ \Phi (a,c,x)$ \cite{3a} we obtained for $G(u)$ the next expression (up to
the first order in $\mu $)\ 
\begin{equation}
G(u)\backsimeq \frac{(2+u)(1+\mu )+\mu (1+u)^{2}}{3+7\mu }.  \label{q11}
\end{equation}%
The relation Eq. \ref{q11} implies that $G(u)$ tends to $G_{0}(u)=\frac{2+u}{%
3}$ when $\mu $ tends to zero. The first two moments of the distribution $%
\rho (n)$ are: $\overline{n}=\frac{1}{3}$ and $\sigma =\frac{2}{9}$. The
relative fluctuation of $n=\frac{\sqrt{\sigma }}{\overline{n}}$ is equal to $%
\sqrt{2\text{ }}$ in this case. We see that in contrast to classical
situation $\overline{n}\neq 0$ when $\mu $ tends to zero moreover
fluctuations of occupation number turn out to be large and very essential.

It is necessary to note that in classical system Eq. \ref{q4} in addition to
variable $\left\vert z\right\vert ^{2}$ we have phase variable $\varphi $
that satisfies to equation $\frac{d\varphi }{dt}=\omega $. However phase
dynamics determines only velocity along limit cycle but does not influence
on stationary state itself. In principle we can turn $\omega $ to zero
(which implies that $H=0$) and all foregoing results do not change. It is
the point which explains why the Lindblad equation can be successfully
applied for statistical description of classical open systems with integer
variables. The reason is that in the case when $H=0$ in Eq. \ref{q4}, 
h{\hskip-.2em}\llap{\protect\rule[1.1ex]{.325em}{.1ex}}{\hskip.2em}
may be eliminated from the Lindblad equation and wave properties of quantum
system turn out to be irrelevant. After this crucial remark we can apply our
approach to different classical systems with integer variables in particular
population dynamics models.

\section{Statistical description of population dynamics models.}

We begin our consideration with well known Lotka-Volterra model (LVM) (see 
\cite{4a}) describing "the interaction" between two populations: prays and
predators (for example hares and lynxes) living in the same territory. Let $%
n_{1}(t)$ and $n_{2}(t)$ the current numbers of preys and predators
correspondently. Then input dynamical equations of the LVM can be written as%
\begin{eqnarray}
\frac{dn_{1}}{dt} &=&n_{1}\left( a-2n_{2}\right) ,  \label{q12} \\
\frac{dn_{2}}{dt} &=&-n_{2}\left( b-2n_{1}\right) ,  \notag
\end{eqnarray}%
where coefficients $a$, $b$ are positive and have clear ecological meaning.
Note that for convenience we have chosen time unit thus to put coefficient
of the term $n_{1}n_{2}$ in Eq. \ref{q12} equal to 2. Now let us demonstrate
that system Eq. \ref{q12} can be represented in FAQ. For this purpose we
introduce two auxiliary complex variables $z_{1}$ and $z_{2}$ so that $\
n_{1}=\left\vert z_{1}\right\vert ^{2}$and $n_{2}=\left\vert
z_{2}\right\vert ^{2}$. Let us assume that evolution $z_{1}$, $z_{2}$ in
time is governed by the following system of equations:%
\begin{eqnarray}
\ \frac{dz_{1}}{dt} &=&\lambda _{1}^{2}z_{1}-z_{1}\left\vert
z_{2}\right\vert ^{2},  \label{q13} \\
\frac{dz_{2}}{dt} &=&-\lambda _{2}^{2}z_{2}+z_{2}\left\vert z_{1}\right\vert
^{2}.  \notag
\end{eqnarray}%
It is easy to see that system Eq. \ref{q13} implies the following equations
for $n_{1}$, $n_{2}$%
\begin{eqnarray}
\frac{dn_{1}}{dt} &=&2\lambda _{1}^{2}n_{1}-2n_{1}n_{2},  \label{q14} \\
\frac{dn_{2}}{dt} &=&-2\lambda _{2}^{2}n_{2}+2n_{1}n_{2},  \notag
\end{eqnarray}%
which is exactly coincides with Eq. \ref{q12} if one lets $a=2\lambda
_{1}^{2}$, $b=2\lambda _{2}^{2}$.

Now if we introduce three functions $R_{1}=\lambda _{1}z_{1}^{\ast }$, $%
R_{2}=\lambda _{2}z_{2}$, and $R_{3}=z_{1}z_{2}^{\ast }$ it is easy to
verify that Eq. \ref{q13} can be represented in the form\ 
\begin{eqnarray}
\frac{dz_{1}}{dt} &=&\sum\limits_{i=1}^{3}\left( \overline{R_{i}}\frac{%
\partial R_{i}}{\partial z_{1}^{\ast }}-R_{i}\frac{\partial \overline{R_{i}}%
}{\partial z_{1}^{\ast }}\right) ,  \label{q15} \\
\frac{dz_{2}}{dt} &=&\sum\limits_{i=1}^{3}\left( \overline{R_{i}}\ \frac{%
\partial R_{i}}{\partial z_{2}^{\ast }}-R_{i}\frac{\partial \overline{R_{i}}%
}{\partial z_{2}^{\ast }}\right) .  \notag
\end{eqnarray}%
Using the foregoing recipe of quantization we maintain that description of
LVM which takes into account the discreteness of variables $n_{1}$\ and $%
n_{2}$\ and their time fluctuations is given by the next master\ equation%
\begin{equation}
\frac{d\rho }{dt}=\sum\limits_{i=1}^{\ 3}\left[ \widehat{R}_{i}\rho ,%
\widehat{R}_{i}^{+}\right] +h.c.,  \label{q16}
\end{equation}%
where $\widehat{R}_{1}=\lambda _{1}\widehat{a}_{1}^{+}$, $\widehat{R}%
_{2}=\lambda _{2}\widehat{a}_{2}$\ and $\widehat{R}_{3}=\widehat{a}_{1}%
\widehat{a}_{2}^{+}$. Again we will interesting in only the solutions of Eq. %
\ref{q16} which have the form $\widehat{\rho }=\sum\limits_{n_{1,n_{2}}}%
\left\vert n_{1}n_{2}\right\rangle \rho _{n_{1,}n_{2}}\left\langle
n_{1}n_{2}\right\vert $. In this case for the coefficients of the expansion $%
\rho _{n_{1}n_{2}}$ we obtain the next general equation 
\begin{widetext}
\begin{equation}
\frac{d\rho _{n_{1}n_{2}}}{dt}=2\lambda _{1}^{2}\left[ n_{1}\rho
_{n_{1}-1,n_{2}}-\left( n_{1}+1\right) \rho _{n_{1}n_{2}}\right] +2\lambda
_{2}^{2}\left[ \left( n_{2}+1\right) \rho _{n_{1,}n_{2}+1}-n_{2}\rho
_{n_{1}n_{2}}\right] +2\left[ \left( n_{1}+1\right) n_{2}\rho
_{n_{1}+1,n_{2}-1}-\left( n_{2}+1\right) n_{1}\rho _{n_{1}n_{2}}\right] 
\label{q17}
\end{equation}
\end{widetext}.\ It is convenient to introduce the generating function $%
G(u,v,t)=\sum\limits_{n_{1}n_{2}}\rho _{n_{1}n_{2}}\cdot u^{n_{1}}\cdot
v^{n_{2}}.$Then after the simple algebra we find that Eq. \ref{q17} implies
the next equation for $G(u,v,t)$:%
\begin{equation}
\frac{\partial G}{\partial t}=2\lambda _{1}^{2}(u-1)\frac{\partial }{%
\partial u}(uG)+2\lambda _{2}^{2}(1-v)\frac{\partial G}{\partial v}+2(v-u)%
\frac{\partial ^{2}}{\partial u\partial v}(vG).  \label{q18}
\end{equation}%
The Eq. \ref{q17} and Eq. \ref{q18} in principle give us all the necessary
information about statistical behaviour of LVM. Now let us consider concrete
results following from these equations. Note that even in stationary case it
is difficult to find exact analytical solutions of Eq. \ref{q18} for
arbitrary $\lambda _{1}$and $\lambda _{2}$. But in the special case when $%
\lambda _{2}^{2}=1+\lambda _{1}^{2}$ such solution easily can be found and
have the form $G(u,v)=\frac{\left( 1-\varkappa \right) ^{2}}{\left(
1-\varkappa u\right) \left( 1-\varkappa v\right) }$ where \ $\varkappa =%
\frac{\lambda _{1}^{2}}{1+\lambda _{1}^{2}}$. Since $G(u,v)=g(u)\cdot g(v)$
it is clear that $n_{1}$and $n_{2}$ are independent variables in this case
and $\overline{n_{1}}=\overline{n_{2}}=\frac{\varkappa }{1-\varkappa }%
=\lambda _{1}^{2}.$The dispersion $\sigma _{1}^{2}$ in this case is equal $\ 
$to $\lambda _{1}^{2}+\lambda _{1}^{4}$ and if we calculate relative
fluctuation \ of $n_{1}$ in stationary state we get the result: $\delta
n_{1}=\frac{\sqrt{\sigma _{1}}}{\overline{n_{1}}}=\sqrt{1+\frac{1}{\lambda
_{1}^{2}}}$. Thereby we see this quantity is not small and can be easily
measured. More detailed analysis of statistical behaviour of LVM following
from Eq. \ref{q17} for arbitrary $\lambda _{1\text{ }}$and $\lambda _{2}$
will be carried out elsewhere. Here we show only that using Eq. \ref{q18}
one can easy\ obtain a collection of explicit relations connecting different
moments of distribution $\rho (n_{1},n_{2})$. For example if we
differentiate stationary Eq. \ref{q18} with respect to $u$ and after that
put $u=v=1$ we obtain the simple relation between moments of first and
second order which reads as:%
\begin{equation}
\lambda _{1}^{2}\left( 1+\overline{n_{1}}\right) -n_{1}-\overline{n_{1}n_{2}}%
=0.  \label{19a}
\end{equation}%
In a similar manner by differentiating of Eq. \ref{q18} with respect to v we
obtain the second relation:%
\begin{equation}
-\lambda _{2}^{2}\overline{n_{2}}\ +\overline{n_{1}}+\overline{n_{1}n_{2}}%
=0\   \label{19b}
\end{equation}%
Relations Eq. \ref{19a} and Eq. \ref{19b} imply the helpful equation
connecting $\overline{n_{1}}$\ and $\overline{n_{2}}$\ namely $\frac{%
\overline{n_{2}}}{1+\overline{n_{1}}}=\frac{\lambda _{1}^{2}}{\lambda
_{2}^{2}}$. It is worth to note that in classical LVM Eq. \ref{q14} similar
relation exists: $\frac{\overline{n_{2}}}{\overline{n_{1}}}=\frac{\lambda
_{1}^{2}}{\lambda _{2}^{2}}$, so we conclude that when $\overline{n_{1}}\gg 1
$, $\overline{n_{2}}\gg 1$ the results of statistical description completely
coincide with pure dynamical consideration. But in the case of small numbers 
$n_{1}$, $n_{2}$ \ the difference between them may be essential. To
demonstrate this distinction and also to compare the approach proposed in
the present paper with usual Markovian description of such systems proposed
in well known article of Nicolis and Prigogine \cite{5a} and expanded in
their later book \cite{6a} it is appropriate to consider the truncated case
of LVM, with $\lambda _{1}=\lambda _{2}=0$. It is obvious that total number
of individuals N$=n_{1}+n_{2}$ in this model will be constant .This fact
greatly simplifies finding and analysis of solutions Eq. \ref{q18} which in
this case takes the form:

\begin{equation}
\frac{\partial G}{\partial t}=(v-u)\frac{\partial ^{2}}{\partial u\partial v}%
(vG)\   \label{q20}
\end{equation}%
It is easy to see that Eq. \ref{q20} has solutions for any integer N in the
form of homogeneous polynomial in $u$ and $v$ of degree N , namely $G_{N%
\text{ }}(u,v,t)=\sum\limits_{k}A_{k}\left( t\right) u^{k}v^{N-k}$ where
coefficients $A_{k}$ satisfy the normalization condition $%
\sum\limits_{k}A_{k}=1$. Thereby Eq. \ref{q20} is reduced to the linear
system of equations of N+1order for coefficients $A_{k}$ of the form$\ \frac{%
dA_{k}}{dt}=L_{km}A_{m}$, where matrix elements $L_{km}$ can directly be
found from Eq. \ref{q20} for any N. For example in simplest case when N=2
the matrix $L_{km\text{ }}$has the form$%
\begin{pmatrix}
-2 & 0 & 0 \\ 
2 & -2 & 0 \\ 
0 & 2 & 0%
\end{pmatrix}%
$.

Let us consider now this case more detail. Let $G_{2}\left( u,v,t\right)
=au^{2}+buv+cv^{2}$ is generating function of the model. Then Eq. \ref{q20}
implies the next system of equations for evolution of coefficients $a,b,c$.

\begin{equation}
\frac{da}{dt}=-2a,\frac{db}{dt}=2a-2b,\frac{dc}{dt}=2b.  \label{q21}
\end{equation}%
Together with normalization condition $a+b+c=1$ system Eq. \ref{q21} allows
one to give statistical description of the model at any time. In particular
Eq. \ref{q21} implies that when t tends to infinity $\overline{n_{1}}$ tends
to zero and $\overline{n_{2\text{ }}}$ tends to 2.This completely agrees
with solutions of dynamical equations in this case. From the other hand let
us consider now the equation for generating function of this model \
obtained by Nicolis and Prigogine (see eq .10.66 in their book \cite{6a})
which in our notation has the form

\begin{equation}
\frac{dG}{dt}=(v-u)v\frac{\partial ^{2}G}{\partial u\partial v}.  \label{q22}
\end{equation}%
In the case $N=2$ Eq. \ref{q22} implies the system of equations for
coefficients of expansion $G_{2}=au^{2}+buv+cv^{2}$ which differs from \ref%
{q21} namely

\begin{equation}
\frac{da}{dt}=0,\frac{db}{dt}=-b,\frac{dc}{dt}=b.  \label{q23}
\end{equation}%
Eq. (\ref{q23}) imply that when t tends to infinity both quantities \ $%
\overline{n_{1}}$, $\overline{n_{2}}$ tend to nonzero values depending on
initial conditions what obviously does not agree with dynamical equations.
But when $N\gg 1$ it is easy to see that asymptotic behaviour solutions
obtained from equations Eq. (\ref{q20}) and Eq. (\ref{q22}) are virtually
identical. Thus in this example we see the benefits of approach proposed
over the standard methods which do not allow to consider explicitly the
discreteness of variables of the problem.

Finally in the last part of the paper using one concrete model of the
population dynamics we want to demonstrate that approach proposed let one to
give statistical description of the systems for which dynamical description
in the framework of mean field theory looks as oversimplified. For this
purpose we consider two "competing" kins of cannibals eating each other so
that voracity of individuals in both kins is assumed to be distinctive. As
one knows cannibalism is widespread \ in living nature and plays important
role in evolution processes \cite{7a}. Besides many species possess special
mechanisms which let them to recognize relatives and to avoid of their
eating \cite{8a}. Let us assume that mutual eating is the major factor \ of
changes of the number of individuals in both kins. Then evolution of the
number of individuals $n_{1}$and $n_{2}$ in such model can be represented of
simple system of equation of the form\ 
\begin{eqnarray}
\frac{dn_{1}}{dt} &=&an_{1}n_{2}-bn_{1}n_{2},  \label{q24} \\
\frac{dn_{2}}{dt} &=&-an_{1}n_{2}+bn_{1}n_{2}.  \notag
\end{eqnarray}%
It is easy to see that total number of individuals in this model N$%
=n_{1}+n_{2}$ conserves. But dynamical description of the system with the
help of Eq. (\ref{q24}) seems to be oversimplified. In particular it implies
that for any N when $a>b$ \ and t tends to infinity $n_{1}$tends to N , and $%
n_{2}$ tends to zero. Now following the spirit of our method we will
describe this system by the help of two operators $\widehat{R_{1}}=\lambda
_{1}\widehat{a_{1}}\widehat{a_{2}^{+}}$ and $\widehat{R_{2}}=\lambda _{2}%
\widehat{a_{1}^{+}}\widehat{a_{2}}$. Show that such statistical version is
completely consistent. Actually acting as in previous examples we can write
the master equation for the density matrix of the system \ as

\begin{equation}
\ \frac{d\widehat{\rho }}{dt}=\left[ \widehat{R_{1}}\widehat{\rho ,}\widehat{%
R_{1}^{+}}\right] +\left[ \widehat{R_{2}}\widehat{\rho },\widehat{R_{2}^{+}}%
\right] +h.c..  \label{q25}
\end{equation}%
If again we are interesting in by solutions of the Eq. (\ref{q25}) of the
form $\widehat{\rho }=\sum\limits_{n_{1}n_{2}}\left\vert
n_{1}n_{2}\right\rangle \rho _{n_{n}n_{2}}\left\langle n_{1}n_{2}\right\vert 
$ then for the generating function of the problem $G(u,v,t)=\sum%
\limits_{n_{1}n_{2}}\rho _{n_{1}n_{2}}u^{n_{1}}v^{n_{2}}$ we obtain the
equation

\begin{equation}
\ \frac{dG}{dt}=a\left( u-v\right) \frac{d^{2}}{dudv}\left( uG\right)
+b\left( v-u\right) \frac{d^{2}}{dudv}\left( vG\right) ,  \label{q26}
\end{equation}%
where $a=2\lambda _{2}^{2}$ \ and $b=2\lambda _{1}^{2}$. The equation Eq. (%
\ref{q26}) implies that generating function of stationary state of this
model in the case when total number of individuals is equal to N can be
represented in the form 
\begin{widetext}
\begin{equation}
G_{st}\left( u,v\right) =C_{N}\left[ \frac{\left( bv\right) ^{N+1}-\left(
au\right) ^{N+1}}{bv-au}\right] =C_{N}\left[ \left( bv\right) ^{N}+\left(
bv\right) ^{N-1}\left( au\right) +...\right]   \label{q27}
\end{equation}
\end{widetext}where $C_{N}$ is the normalization factor, $C_{N}=\left(
b^{N}+ab^{N-1}+...\right) ^{-1}$ Having in hands expression Eq. (\ref{q27})
for the generation function one can find all statistical characteristics of
the model for any N. In particular for N=2 for the average values of
individuals in both kins we obtain $\overline{n_{1}}=\frac{2a^{2}+ab}{%
b^{2}+ab+a^{2}}$, and $\ \overline{n_{2}}=\frac{2b^{2}+ab}{b^{2}+ab+a^{2}}$.
Now we want to show that when N tends to infinity the properties of our
statistical model will be similar to ones of the dynamical model Eq. (\ref%
{q24}).Let us assume that $\ b>a$ and let $\ \ \ \ \ \ \ \varkappa =\frac{a}{%
b},\left( \varkappa <1\right) .$ Let us calculate now the ratio $\frac{%
\overline{n_{1}}}{\overline{n_{2}}}$ in this model. Using Eq. (\ref{q27}) we
obtain 
\begin{widetext}
\begin{equation}
\frac{\overline{n_{1}}}{\overleftarrow{n_{2}}}=\frac{Na^{N}+\left(
N-1\right) a^{N-1}b+...}{Nb^{N}+\left( N-1\right) b^{N-1}a+...}=\frac{%
N\varkappa ^{N}+\left( N-1\right) \varkappa ^{N-1}+...\varkappa }{N\text{ }%
+\left( N-1\right) \varkappa +....\varkappa ^{N-1}}  \label{q28}
\end{equation}
\end{widetext}. Expression Eq. (\ref{q28}) can be represented in the next
convenient form 
\begin{equation}
\frac{\overline{n_{1}}}{\overline{n_{2}}}=\frac{\varkappa +\varkappa ^{2}%
\frac{\partial Lnf_{N}\left( \varkappa \right) }{\partial \varkappa }}{%
N-\varkappa \frac{\partial Lnf_{N}\left( \varkappa \right) }{\partial
\varkappa }},  \label{q29}
\end{equation}%
where $f_{N}\left( \varkappa \right) =\frac{1-\varkappa ^{N}}{1-\varkappa }$%
. It is easy to see that $\frac{\partial \left( Lnf_{N}\right) }{\partial
\varkappa }=\frac{\left( N-1\right) \varkappa ^{N}-N\varkappa ^{N-1}+1}{%
\left( 1-\varkappa \right) \left( 1-\varkappa ^{N}\right) }$ . Obviously
this expression tends to $\frac{1}{1-\varkappa }$ when $\varkappa <1$ and N
tends to infinity .Thus Eq. (\ref{q29}) implies that ratio $\frac{\overline{%
n_{1}}}{\overline{n_{2}}}$ tends to zero but this case is realized only for
infinite population. From the other hand for any finite population of
cannibals all its statistical properties can be well described by the
approach proposed in our paper. In conclusion let us briefly summing up our
consideration. We propose the statistical method of describing of various
systems in physics,chemistry,ecology \ whose states can be represented by
integers. Although the method proposed is based on quantum Lindblad equation
nevertheless we have shown that it can be successfully used also for
description of classical open systems with integer variables (at least in
the cases when "Hamiltonian" of the open system is equal to zero) .All
examples considered in present paper demonstrate that approach proposed
results in consistent conclusions and allows one in several cases to
eliminate essential inaccuracies of preceding considerations.

\section{Acknowledgement}

The author acknowledges I.V. Krive and S.N. Dolya for the discussions of the
results of the paper and valuable comments.

\end{document}